\documentclass[aps,prd,superscriptaddress,floatfix,nofootinbib,notitlepage,letterpaper]{revtex4}

\usepackage{graphicx}
\usepackage{amsmath}
\usepackage{aas_macros}
\usepackage{xcolor}

\usepackage[sort&compress]{natbib}
\usepackage{hyperref}
\newcommand{\irf}[1]{\texttt{#1}}

\newcommand{\Fermi}{{\textit{Fermi}}}
\newcommand{\fermi}{\Fermi}

\usepackage{color} 

\ifdefined\bwfigures
\else
\fi

\newcommand\fdg{\mbox{$~.\!\!^\circ$}} 
\newcommand\dg{\mbox{$~\!\!^\circ$}}

\begin{document}

\title{Search for Gamma-ray Emission from Dark Matter Annihilation in the Small Magellanic Cloud with the \fermi\ Large Area Telescope}

\date{\today}

\author{R.~Caputo}
\email{rcaputo@ucsc.edu}
\affiliation{Santa Cruz Institute for Particle Physics, Department of Physics and Department of Astronomy and Astrophysics, University of California at Santa Cruz, Santa Cruz, CA 95064, USA}
\author{M.~R.~Buckley}
\email{mbuckley@physics.rutgers.edu}
\affiliation{Rutgers, The State University of New Jersey, Department of Physics and Astronomy, Piscataway, NJ 08854, USA}
\author{P.~Martin}
\affiliation{CNRS, IRAP, F-31028 Toulouse cedex 4, France}
\author{E.~Charles}
\email{echarles@slac.stanford.edu}
\affiliation{W. W. Hansen Experimental Physics Laboratory, Kavli Institute for Particle Astrophysics and Cosmology, Department of Physics and SLAC National Accelerator Laboratory, Stanford University, Stanford, CA 94305, USA}
\author{A.~M.~Brooks}
\affiliation{Rutgers, The State University of New Jersey, Department of Physics and Astronomy, Piscataway, NJ 08854, USA}
\author{A.~Drlica-Wagner}
\affiliation{Center for Particle Astrophysics, Fermi National Accelerator Laboratory, Batavia, IL 60510, USA}
\author{J.~Gaskins}
\affiliation{GRAPPA, University of Amsterdam, Science Park 904, 1098XH Amsterdam, Netherlands}
\author{M.~Wood}
\affiliation{W. W. Hansen Experimental Physics Laboratory, Kavli Institute for Particle Astrophysics and Cosmology, Department of Physics and SLAC National Accelerator Laboratory, Stanford University, Stanford, CA 94305, USA}

\begin{abstract}
The Small Magellanic Cloud (SMC) is the second-largest satellite galaxy of the Milky Way and is only 60~kpc away.
 As a nearby, massive, and dense object with relatively low astrophysical backgrounds, it is a natural target for dark matter 
indirect detection searches. In this work, we use six years of {\tt Pass 8} data from the {\it Fermi} Large Area Telescope to search 
for gamma-ray signals of dark matter annihilation in the SMC. 
Using data-driven fits to the gamma-ray backgrounds, and a combination of N-body simulations and direct measurements 
of rotation curves to estimate the SMC DM density profile, we found that the SMC was well described 
by standard astrophysical sources, and no signal from dark matter annihilation was detected.  
We set conservative upper limits on the dark matter annihilation cross section. These constraints are in agreement with 
stronger constraints set by searches in the Large Magellanic Cloud and approach the canonical thermal relic cross section at 
dark matter masses lower than 10 GeV in the $b\bar{b}$ and $\tau^+\tau^-$ channels.

\end{abstract}
\maketitle

\section{INTRODUCTION}\label{sec:intro}

It has been clear for many decades that the observed Universe includes a significant component of matter which does not interact like any known field in the Standard Model of particle physics. 
Though solid observational evidence exists for the gravitational influence of this dark matter (DM) from the earliest 
moments of the Universe's history to the present day \cite{Zwicky:1933gu,Rubin:1980zd,Olive:2003iq,Ade:2013zuv}, 
no direct measurements have been made of the particle nature of this mysterious substance. 

Though by no means the only possibility, a theoretically well motivated class of DM models has interactions between itself and the Standard Model that are 
approximately as strong as the weak nuclear force, and a mass of similar scale ($\sim$10--1000  GeV). Such weakly interacting massive particles (WIMPs) 
would generically attain the observed DM density after thermal freeze-out in the early Universe. The canonical ``WIMP'' is a $\sim$100 GeV particle 
interacting through the $SU(2)_L$ weak force, though many other candidates have been proposed with a wide range of masses and interaction strengths \cite{Feng:2008ya}.

These models provide a useful benchmark for DM searches designed to look for the present-day pair annihilation (or decay) of DM particles in regions of high density of DM. 
A thermally-averaged annihilation cross section of $\langle \sigma v\rangle \sim 3 \times 10^{-26}$~cm$^3$/s results in approximately the correct WIMP relic density, and so 
experiments capable of seeing the present-day annihilation of DM with cross sections near this value have the sensitivity to either confirm
or exclude a large number of theoretically interesting models.

Within the paradigm of annihilating DM, there are many possibilities for the annihilation channel. Of particular interest is annihilation resulting in gamma rays, as this signature is 
more easily distinguished from other astrophysical sources. However this annihilation channel is suppressed; thus, searches for this signature are challenging. 
In addition to the direct annihilation to pairs of photons, if DM annihilates into pairs of other Standard Model particles, 
the resulting hadronization and/or decay will result in a continuum of gamma rays observable from Earth with an energy distribution that extends up to the rest mass of the DM particle. 
Gamma rays are also relatively unaffected by the intervening medium and arrive at the Earth unscattered and unattenuated (at least in the local Universe), which  
allows the emission to be tracked. Gamma-ray observations together with separate information or assumptions about the distribution of DM in the region under study 
and models for the hadronization then allow measurement, or determination of upper limits for, the annihilation cross section. 

With this motivation in mind, the gamma-ray data set compiled by the Large Area Telescope (LAT) carried by the {\it Fermi Gamma-ray Space Telescope} 
({\it Fermi}-LAT) is of great interest. At the present time, the {\it Fermi}-LAT is one of the most sensitive instruments to DM with weak-scale mass and 
cross section annihilating into gamma rays. Analysis of the LAT gamma-ray data can place strong limits on---or discover---DM  
annihilation with cross sections near the canonical thermal value into a wide variety of Standard Model particles~\cite{Atwood:2009ez}. 

A large number of DM searches have been performed using the {\it Fermi}-LAT data; as annihilation rates are proportional to the square of the DM density, 
lower annihilation cross sections can be probed by targeting regions of the sky with the greatest densities of DM, such as the center of the Milky Way~\cite{Abazajian:2012pn,Abazajian:2014fta,Boyarsky:2010dr,Daylan:2014rsa,Goodenough:2009gk,Gordon:2013vta,Hooper:2010mq,Hooper:2011ti,Hooper:2012sr,Hooper:2013rwa,Huang:2013pda, TheFermi-LAT:2015kwa}, 
satellite dwarf spheroidal galaxies of the Milky Way~\cite{Ackermann:2011wa,Ackermann:2013yva,GeringerSameth:2011iw,Geringer-Sameth:2014qqa,Ackermann:2015zua,Geringer-Sameth:2015lua,Drlica-Wagner:2015xua}, 
unresolved halo substructure~\cite{Ackermann:2012nb,Belikov:2011pu,Berlin:2013dva,Buckley:2010vg}, 
galaxy clusters~\cite{Ackermann:2010rg,Dugger:2010ys}, 
and the Large Magellanic Cloud~\cite{Buckley:2015doa}. 

The LAT observations of the Galactic center indicate that the region is brighter than expected from standard models for Galactic diffuse emission in 
the few-GeV range, and the spatial distribution is broadly consistent with our expectations for a DM signal. However, previously unconsidered astrophysical 
backgrounds could match the observed morphology 
and spectrum~\cite{Abazajian:2010zy,Wharton:2011dv,Hooper:2013nhl,Carlson:2014cwa,Petrovic:2014xra,Petrovic:2014uda,2013MNRAS.436.2461M}, 
and the true source of the gamma rays remains a subject of much debate.

Considering both the broad interest in indirect searches for DM, and the current questions raised by the Galactic center excess, 
it is important to identify new high-density targets for DM annihilation indirect searches. Here we apply the techniques developed in 
the search for DM in the Large Magellanic Cloud (LMC)~\cite{Buckley:2015doa}, to a similar analysis of the Small Magellanic Cloud (SMC). 
The SMC is a satellite galaxy of the Milky Way, approximately 60~kpc away and with a DM mass of $\sim$10$^{10}~\mathrm{M}_\odot$ within a radius 
of $\sim$3 kpc \cite{Bekki:2008db}. The SMC is in a complicated orbit with its larger cousin the LMC~\cite{Besla:2010ws}, and the combined 
Magellanic system appears to be on its first approach to the Milky Way~\cite{Besla2007,Kallivayalil:2013xb}. 
Therefore, the Clouds are not expected to have been substantially tidally stripped of DM by our Galaxy. Though the SMC DM profile may be 
affected by its interactions with the LMC, direct measurements of the rotation curve of the SMC indicate significant amounts of DM remain bound to 
the galaxy itself. As we shall demonstrate in this paper, the combination of rotation curves and comparison with cosmological simulations 
of galaxies of the same size that include baryonic physics indicates that at minimum the amount of DM present in the SMC would result in a DM 
annihilation signal as large as the brightest dwarf galaxies, though somewhat dimmer than the LMC itself. This lower signal is offset by the lower 
gamma-ray background in the SMC compared to the LMC, as modeled using {\it Fermi}-LAT data~\cite{2010A&A}, and as a result the SMC is 
an attractive target for DM indirect detection searches.

In Section~\ref{sec:dm}, we describe the DM distribution in the SMC and how it relates to searches for indirect signals of DM annihilation. 
In Section~\ref{sec:smc}, we discuss the {\it Fermi}-LAT instrument, the method of modeling the SMC as a gamma-ray source, 
and the data set and background models used for the DM analysis. 
The analysis techniques and the resulting bounds are shown in Sections~\ref{sec:analysis} and ~\ref{sec:results}, and we conclude in Section~\ref{sec:conclusion}. 

\section{SMC DARK MATTER DISTRIBUTION}\label{sec:dm}

The flux spectrum $d\phi/dE_\gamma$ of gamma rays from any distribution of DM depends on a number of quantities, 
which can be factored into astrophysics- and particle physics-dependent terms~\cite{Ullio:2002pj}:

\begin{equation}
\frac{d\phi}{dE_\gamma} = \left(\frac{x \langle \sigma v\rangle}{8\pi} \frac{dN_\gamma}{dE_\gamma} \frac{1}{m_\chi^2} \right)\left(\int_{\Delta\Omega} d\Omega	\int_{\rm l.o.s.}d\ell \rho^2_\chi(\vec{\ell}) \right). \label{eq:dmflux}
\end{equation}
The quantities in the first parentheses are the annihilation cross section-speed product averaged over the velocity distribution of the DM particles $\langle \sigma v\rangle$, 
the differential yield of gamma rays from a single DM annihilation $dN_\gamma/dE_\gamma$, the mass of the DM particle $m_\chi$, and a normalization factor $x$ which is unity if 
the DM is its own antiparticle and $1/2$ if it is not. All of these depend on the unknown microphysics in the dark sector. The typical approach of DM indirect detection searches, as 
we will follow here, is to place an upper bound (if no excess is observed) on $\langle \sigma v\rangle$ as a function of mass $m_\chi$, assuming a particular spectrum $dN_\gamma/dE_\gamma$ 
and value for $x$, the result of making a particular choice for the DM annihilation channel. In this paper, we assume $x=1$ and consider the final states $b\bar{b}$ and $\tau^+ \tau^-$, 
which have been of particular interest given the Galactic center excess.
Other sets of Standard Model final states are possible, but have sufficient similarity to the channels selected that bounds can reasonably be extrapolated. In this work, we calculate the spectrum 
$dN_\gamma/dE_\gamma$ for each final state and DM mass choice using code available as part of the {\it Fermi}-LAT 
{\em ScienceTools}.\footnote{The {\tt DMFitFunction} spectral model described at 
\url{http://fermi.gsfc.nasa.gov/ssc/data/analysis/documentation/Cicerone/Cicerone_Likelihood/Model_Selection.html}, see also Ref.~\cite{Jeltema:2008hf}.}  
Note that our implementation does not include electroweak corrections~\cite{2007PhRvD..76f3516K,2002PhRvL..89q1802B,2009PhRvD..80l3533K,2010PhRvD..82d3512C,Ciafaloni:2010ti}. 
Such corrections can be important for heavy DM ($m_\chi \gtrsim 1$~TeV); in any case, they would increase the resulting flux and 
thus strengthen the resulting bounds~\cite{Ciafaloni:2010ti,Cirelli:2010xx, Ackermann:2015tah}.  

In order to describe experimental results in terms of the particle physics quantities in Eq.~\eqref{eq:dmflux}, the astrophysical quantities in the second set of 
parentheses must be known. This quantity, the integral of the square of the DM density along the line of sight and over a solid angle $\Delta\Omega$ 
corresponding to the region under study, is known as the $J$-factor, and encapsulates the dependence of an indirect detection search on the distribution of DM in the search target. 
As the $J$-factor depends on the density squared for annihilating DM and implicitly on inverse distance squared, targeting nearby overdensities of DM yields 
larger values of the $J$-factor and thus results in searches that probe smaller annihilation cross sections $\langle \sigma v\rangle$. 
In order to extract results from an analysis of the SMC that can be compared with indirect detection searches targeting other astrophysical objects, 
we must determine the DM density distribution of the SMC, and from this calculate the $J$-factor.

Our fit to the SMC density profile as a function of radius $r$ is parametrized in terms of a generalized Navarro-Frenk-White (NFW) 
profile~\cite{1990ApJ...356..359H,REF:1996ApJ...462..563N,1996MNRAS.278..488Z,1998ApJ...502...48K}

\begin{equation}
\rho(r) = \frac{\rho_0}{\left(\frac{r}{r_S}\right)^\gamma \left[ 1+\left(\frac{r}{r_S}\right)^\alpha\right]^{\frac{\beta-\gamma}{\alpha}}} \Theta(r_\text{max}-r), \label{eq:general_nfw}
\end{equation}
where $r_S$ is a scale radius, $r_{\rm max}$ is a maximum radius which we set to $100~$kpc (our results are relatively insensitive to this choice), 
$\rho_0$ is a normalization factor, $\Theta$ is the Heaviside step-function, and $(\alpha,\beta,\gamma)$ control the inner and outer slopes of the profile. 
The classical NFW profile has $(\alpha,\beta,\gamma) = (1,3,1)$. We will determine the best fit for the free parameters $r_S$, $\rho_0$, and 
$(\alpha,\beta,\gamma)$ to the SMC in two ways: first by using observations of the SMC to directly build the rotation curve and from there infer the DM 
in the inner regions, and second by comparing the SMC to similar galaxies drawn from cosmological simulations.

Proper motion data for the LMC indicate that it may be on its first infall into the Milky Way's DM halo~\cite{Besla2007}. The LMC and SMC are likely a pair of dwarf galaxies 
that are being accreted to the Milky Way together.  Both the Magellanic Bridge (H\,{\sc I} gas joining the two galaxies) and the Magellanic Stream (H\,{\sc I} 
gas trailing the orbit of the Clouds) are best explained by tidal interactions between the two galaxies before infall into the Milky Way \cite{Besla:2012fp}.  As the smaller of the two 
galaxies, this may mean that the SMC has been tidally harassed by the LMC, leading to a somewhat complicated structure.  The SMC is elongated along our line of sight, with a 
bar-like body that we may be viewing end-on (see {\it e.g.}~Refs.~\cite{Haschke:2012mz,2015arXiv150206995S,2015AJ....149..179D}). In fact, it has been suggested that the 
stellar bar of the LMC is a remnant of a passage by the SMC through the LMC, leading to a bar that is elevated above the LMC's disk plane \cite{Besla:2012fp}. Despite these 
complications, we demonstrate below that simple DM halo models derived from isolated halos provide a remarkably good fit to the observed rotation curve of the SMC.

Under the assumption of circular orbits, the rotational velocity of a galaxy is a direct measurement of the mass enclosed as a function of radius, 
$v^2_{\rm rot} = GM(<r)/r$. The contribution of DM to the H\,{\sc I} rotation curve of the SMC has been studied in detail by 
Refs.~\cite{Stanimirovic:2003mp,Bekki:2008db}, and we adopt some of their results in this work.  To briefly summarize, Ref.~\cite{Stanimirovic:2003mp} 
fit a tilted ring model to the H\,{\sc I} data, and corrected for asymmetric drift (the velocity dispersion of the SMC is fairly large, contributing as much 
dynamical support as the rotational support at some radii).  A mean inclination of $i = 40\dg \pm 20\dg$ was found.  By fitting an exponential disk with a scale 
height of 1 kpc, the  H\,{\sc I}+He mass of the SMC was found to be 5.6$\times$10$^8$ M$_{\odot}$.  Ref.~\cite{Bekki:2008db} adopted $i = 40\dg$, and explored 
varying stellar mass-to-light ratios ($M/L_V$, where $L_V$ is the $V$-band luminosity of 4.3$\times$10$^8$ $L_{\odot}$). Based on the SMC's derived star formation 
history~\cite{HarrisZaritsky2004}, Ref.~\cite{Bekki:2008db} suggest that $M/L_V$ in the range of 2--4 is reasonable.  They adopt $M/L_V =$ 2.3 as their fiducial model, 
giving a SMC stellar mass of 9.9$\times$10$^8$ M$_{\odot}$ with $r_s = 5.1$ kpc.

\begin{figure}[ht]
\includegraphics[width=0.6\columnwidth]{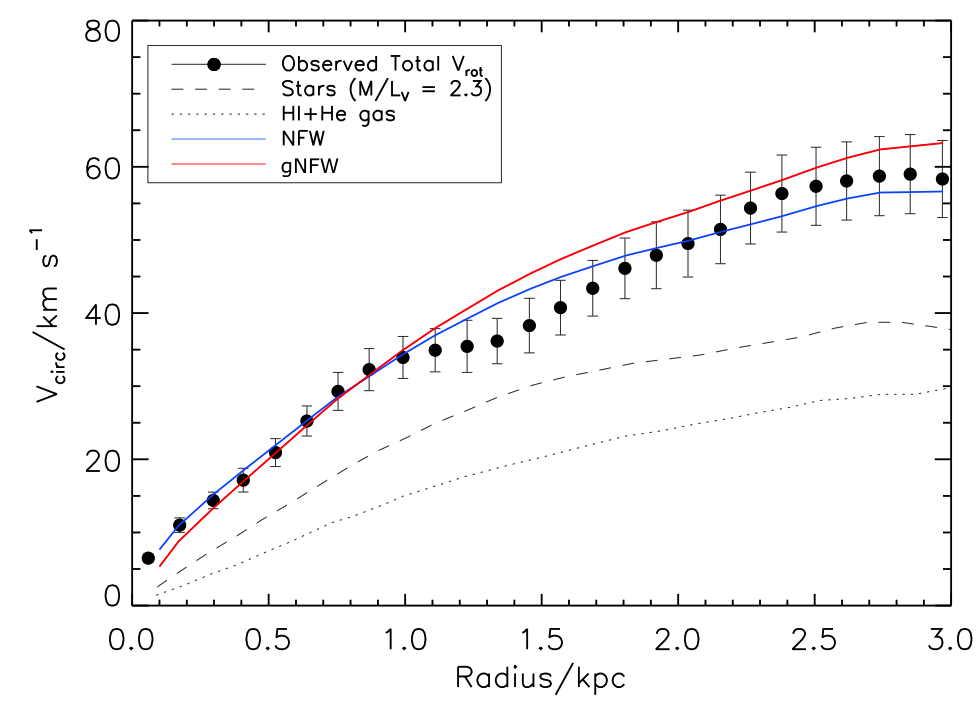}
\caption{Observed H\,{\sc I} rotation curvefor the SMC (black dots) from Refs.~\cite{Bekki:2008db} assuming $M/L_V = 2.3$ and an inclination angle $i = 40\dg$, and 
contributions to rotation curve from stellar mass (dashed line) and H\,{\sc I}+He gas (dotted line). The overall rotation curve including DM contribution of best-fit NFW 
profile from Ref.~\cite{Bekki:2008db} is in blue, and that for the generalized NFW extracted from simulation using results of Ref.~\cite{diCintio2014b} 
is in red. \label{fig:SMCrotcurve}}
\end{figure}

In Figure~\ref{fig:SMCrotcurve} we show the rotation curve data, adopted from Ref.~\cite{Bekki:2008db}.  We assume $i = 40\dg$, and $M_{\rm stellar}=$ 9.9$\times$10$^8$ 
M$_{\odot}$ ({\it i.e.}, $M/L_V =$ 2.3).  The lines show the contribution from the  H\,{\sc I}+He gas, the stellar component, and the model that best fits 
both the inner and outer observed total rotation curve. This model has an NFW profile, and is denoted as the {\tt NFW} model for the remainder of the paper.

As in Ref.~\cite{Buckley:2015doa}, we also use state-of-the-art cosmological simulations to determine the ``typical'' DM density profile for a galaxy with 
the stellar mass of the SMC.  We use a set of simulations that have shown that energetic feedback from stars and supernovae can transform an initially 
steep inner density profile into a shallower profile~\cite{Governato2012, Teyssier2013, diCintio2013}. The degree of transformation is sensitive to the mass 
of stars formed~\cite{Penarrubia:2012bb,diCintio2013}, and the stellar mass is dependent on halo mass~\cite{Behroozi2013, Moster2012}. These simulations 
have been shown to match the observed stellar-to-halo mass relation at redshift $z = 0$ [94].  Ref.~\cite{diCintio2014b} has provided a general 
relation for the generalized NFW parameters ($\alpha$, $\beta$, $\gamma$) as a function of stellar-to-halo mass ratio. Therefore, we can extract a range 
of generalized NFW profiles appropriate for the SMC from simulations.

To find the generalized NFW parameters that best describe the SMC, we searched for simulated galaxies in the available set that have a luminosity in the 
same range as the SMC.  We adopt a stellar mass for the SMC in the range 4--16$\times 10^8 \mathrm{M}_{\odot}$ (by adopting a luminosity of 
$4\times 10^8 \mathrm{L}_{\odot}$ and $1 < M/L < 4$).  Five simulated galaxies were in this stellar mass range, and had $-2.0 < \log_{10}(M_{\rm stellar}/M_{\rm halo}) < -1.7$.  
We averaged the generalized profile from Ref.~\cite{diCintio2013} at these two extremes, and found it to yield a remarkably good fit to the observed rotation curve 
when added to the stellar and H\,{\sc I} contributions, shown by the red line in Figure~\ref{fig:SMCrotcurve}. This model is referred to as the 
{\tt gNFW} model below.

In Table~\ref{tab:profiles}, we summarize the two DM profiles, {\tt NFW} and {\tt gNFW}, which we will use for the remainder of the analysis. 
We explore both profiles because they span a range of density profiles consistent with the observations. The NFW model was shown by Ref.~\cite{Bekki:2008db} 
to be a good fit to the observed rotation curve, and has the expected steep DM density profile that is consistently produced in $N$-body 
simulations of galaxy formation in Cold Dark Matter (CDM).  However, observations and recent simulations that include energy feedback from stars suggest that 
the slopes of the DM density profiles of galaxies with the stellar mass of the SMC should be much shallower than NFW 
(Refs.~\cite{Governato2012, Teyssier2013, diCintio2013,Penarrubia:2012bb}).  
Adopting these two models allows us to explore the range of allowed profiles.
The $J$-factors for each profile were calculated assuming an SMC distance of 60~kpc \cite{Graczyk:2013bpa}. 
We adopt as the center of the SMC the reported stellar kinematic center at J2000 epoch 
$(\alpha,\delta) = (13\fdg19,-72\fdg83)$, though we 
will later scan over the possible central location when performing our fit. As can be seen, the resulting $J$-factors of 
$\log_{10} J/(\text{GeV}^2/\text{cm}^5) \sim$19.1--19.6 are competitive with the best dwarf galaxies, which have 
$\log_{10} J/(\text{GeV}^2/\text{cm}^5) \sim$19--19.5~\cite{Martinez2013,Geringer-Sameth:2014yza}. For comparison, the Galactic center, 
integrated over the inner $1\dg$, has $\log_{10} J/(\text{GeV}^2/\text{cm}^5) \gtrsim$21--24~\cite{Daylan:2014rsa}, and the 
LMC has $\log_{10} J/(\text{GeV}^2/\text{cm}^5) \sim$19.5--20.5~\cite{Buckley:2015doa}. We should emphasize that it is possible 
for the SMC DM to be more concentrated in the inner region, resulting in a significantly larger $J$-factor; however to set conservative lower limits we 
adopt this lower range.

\begin{table}[ht]
\begin{tabular}{ccccccc}
\hline \hline
Profile & $\alpha$ & $\beta$ & $\gamma$ & $r_S$ (kpc) & $\rho_0$ (M$_\odot/\mbox{kpc}^3$) & $J$ (GeV$^2$/cm$^5$) \\ \hline
{\tt NFW} & 1 & 3 & 1 & 5.1 & $4.1\times 10^6$ & $1.13\pm0.01 \times 10^{19}$ \\
{\tt gNFW} & $1.8\pm0.35$ & $2.65\pm0.06$ & $0.69\pm0.14$ & 5 & $7.0 \times 10^6$ & $4.56\pm0.05 \times 10^{19}$ \\ \hline
\end{tabular}
\caption{Summary of the two DM density profiles we adopt for the SMC, including $J$-factor. 
The $J$-factor is computed by integrating to an angular distance of 15$\dg$ from the SMC center; 
however, the majority of the contribution comes from the innermost degrees. For example, using the {\tt gNFW} density profile and integrating to an angular distance of 1$\dg$ 
yields a $J$-factor of $7.98\pm0.01 \times 10^{18}$.
Parameters for the NFW profile were derived from a fit to the observed rotation curve in Ref.~\cite{Bekki:2008db}. 
The gNFW parameters were instead derived from a best fit to simulated DM halos of similar properties to those of the SMC, see Ref.~\cite{diCintio2014b} and text for details.
\label{tab:profiles}}
\end{table}

\section{FERMI-LAT OBSERVATIONS OF THE SMC}\label{sec:smc}

The {\it Fermi}-LAT is a pair-conversion telescope. 
Incoming gamma rays pass through the anti-coincidence detector and convert in the tracker to $e^{+}/e^{-}$ pairs. 
The charged particle direction is reconstructed using the information in the tracker, and the energy is estimated from depositions in the calorimeter.  
Detailed descriptions of the LAT and its performance can be found in dedicated papers~\cite{Atwood:2009ez,2013arXiv1303.3514A}. 

The SMC was detected in gamma rays for the first time by the LAT~\cite{2010A&A}. 
The analysis of 17 months of all-sky observations led to the detection of an extended source $\sim$3$\dg$ in size, 
approximately the angular size of the SMC in various bands, with a significance of about 11$\sigma$.
The emission is steady and has an integrated $>100$~MeV photon flux of $(3.7\pm0.7)\times10^{-8}$ ph/cm$^{2}$/s. 
No obvious spatial correlation of the gamma-ray emission with known components of the SMC was observed, which made
it hard to pinpoint the origin of the emission.The spectrum of the emission was consistent with emission arising from cosmic rays 
interacting with the interstellar medium in the SMC, but a population of high-energy pulsars could also account for a substantial fraction of the signal.

Compared to this early work, about five times more {\it Fermi}-LAT data are now available, and these include improvements in instrument calibration, event 
reconstruction, and background rejection ($i.e.$ the upgrade from {\tt Pass 6} to {\tt Pass 8} data; see Section \ref{sec:fermi}).
We have revisited the study of the gamma-ray emission from the SMC based on this enlarged data set, and the full analysis will be 
presented elsewhere ({\it Fermi}-LAT Collaboration, 2016, in prep.).
We briefly summarize below the results of the analysis, focusing on what is relevant to the present work.
Understanding the methodology and uncertainties associated with the modeling of the gamma-ray emission of the SMC 
defines the limitations of searching for DM signals in this region. 

The analysis to develop the SMC model uses {\tt Pass 8} SOURCE class events in the energy range 200 MeV--300 GeV with a zenith angle cut at 100$\dg$.
The lower energy bound is determined by the worsening angular resolution, and declining acceptance, with decreasing energy.
We apply the standard selection criteria for good time intervals and remove data taken in non-standard operating and 
observing modes. The selection also excludes time periods during bright GRBs and solar flares.
The data set that was used to develop our model for the SMC diffuse emission overlaps with the data set used in the DM search.
The emission model is built from a fitting procedure using a maximum likelihood approach similar to the development of the LMC background 
model~\cite{Buckley:2015doa}.

\subsection{Modeling the SMC}\label{sec:model_smc}

The base model is composed of an isotropic diffuse component, a Galactic diffuse component,\footnote{The diffuse background models are available 
at: \url{http://fermi.gsfc.nasa.gov/ssc/data/access/lat/BackgroundModels.html} as \irf{iso\_P8R2\_SOURCE\_V6\_v06.txt} and \irf{gll\_iem\_v06.fits}.} 
and objects listed in the \textit{Fermi}-LAT Third Source Catalog (3FGL)~\cite{REF:2015.3FGL}. 
Source 3FGL~J0059.0$-$7242e is not included because it corresponds to the SMC, for which we sought a new model.
The position of weak and soft source 3FGL~J0021.6$-$6835 ($(\alpha,\delta)=(5\fdg4, -68\fdg6)$) within the region under study had to be refined because the fit resulted in a pair 
of negative and positive count residuals and an unphysically soft spectrum for the source. 
The new source, referred to as PS2 hereafter, best-fit position is $(\alpha,\delta)=(5\fdg9, -68\fdg3)$. 
In the course of the analysis, an additional point-like source not listed in the 3FGL was found\footnote{The 3FGL catalog is based on 4 years of {\tt P7Rep} 
data, so finding a new source with 6 years of {\tt Pass 8} data is not unexpected.}, between the 
SMC and globular cluster 47~Tuc, at position $(\alpha,\delta)=(10\fdg1, -71\fdg9)$, referred to as PS1 hereafter. It has a TS in the 25--35 range, 
depending on the spatial model adopted for the SMC. Its spectrum can be described by a power law with photon index 1.8. 

The remaining emission coincident with the SMC and not accounted for by the base model is then modeled in several ways.
Representing the SMC as a combination of point sources resulted in too many degrees of freedom for a relatively limited improvement in the maximum likelihood.
Using a 2D Gaussian intensity distribution provides a better likelihood for a smaller number of parameters.
The maximum likelihood 2D Gaussian model is centered on $(\alpha,\delta)=(14\fdg2, -72\fdg8)$ with a size $\sigma = 0\fdg8$.
A combination of 2D Gaussians was also considered as a possible model, but adding a second Gaussian to the one previously described resulted in 
only a negligible improvement of the the maximum likelihood.
A more physically motivated model was tested, the so-called \textit{emissivity model}, similar to what was performed for the LMC analysis~\cite{TheFermi-LAT:2015lxa}.
The emissivity model relies on the assumption that the gamma-ray emission results from cosmic rays interacting with interstellar gas in the SMC.
The aim of this approach is to determine a spatial distribution for the emissivity, which is the gamma-ray luminosity per H-atom per solid angle and depends 
on the actual density and spectrum of cosmic rays threading the gas.
We performed an iterative search including sources with TS $\geq$25 to find the combination of 2D Gaussian emissivity profiles that, after multiplication by the gas column density 
distribution, provides the overall maximum likelihood to the LAT data. 
The maximum likelihood model thus obtained is based on a single 2D Gaussian emissivity profile (compared to 5 for the LMC) 
in addition to the measured distribution of gas in the SMC, centered on $(\alpha,\delta)=(13\fdg2, -72\fdg5)$ with a size $\sigma = 1\fdg4$.
The corresponding log of the maximum likelihood is increased compared to the Gaussian model by about 13, but the significance of that improvement is 
not easily quantified because the models are not nested.

Overall, both spatial models can be considered as two alternatives for the spatial modeling of the gamma-ray emission of the SMC 
and the slightly higher likelihood of the emissivity model should not be taken as a proof that the signal originates predominantly in cosmic rays.
Using either model, the SMC is detected with a significance of nearly 28$\sigma$, with an integrated $>100$~MeV photon flux of $(4.7\pm0.7)\times10^{-8}$ ph/cm$^{2}$/s
(extrapolated from the $>200$~MeV analysis).
The maximum likelihood spectral model, among those tested, is a power law with an exponential cutoff at $8\pm4$ GeV that is significant at the $>4\sigma$ level.

\subsection{Data Selection}\label{sec:fermi}

For this analysis, we use six years of LAT data (2008 August 4 to 2014 August 5) selecting \irf{Pass 8} SOURCE-class events 
in the energy range from 500 MeV to 500 GeV in 24 logarithmic energy bins and with 0$\fdg$1 angular pixelization. 
The data selection used in the DM search is very similar to that used to build the background model described earlier in this section, 
but is shifted to a higher energy range than the selection used to build the background model described in Section~\ref{sec:smc}. We model the 
performance of the LAT using the \irf{P8R2\_SOURCE\_V6} Instrument Response Functions (IRFs).
The lower limit of 500 MeV was chosen to mitigate both the impact of the increasing width of the point-spread function at lower energies and the leakage 
from the Earth's limb (terrestrial gamma rays). 

\begin{figure}[ht]
\includegraphics[width=0.6\columnwidth]{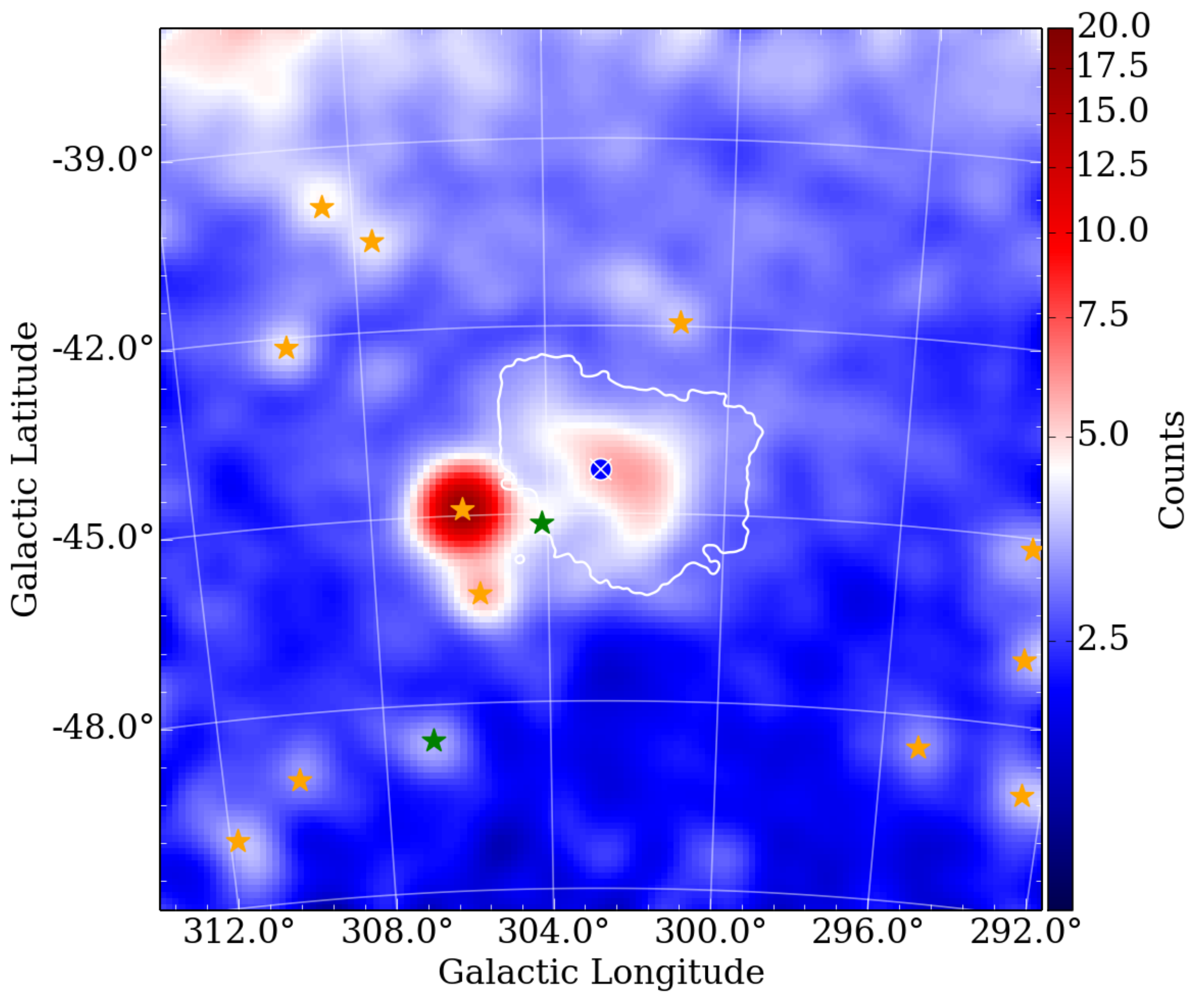}
\caption{Counts map of gamma rays in the SMC region, for 0.5--500~GeV.
The diffuse emission of the SMC from Section~\ref{sec:smc} is outlined in the white solid contour representing 4\% of the maximum.  
The kinematic center of the SMC is the blue circle with the white ``X" through it. 
Orange stars indicate sources in the 3FGL. Green stars indicate the two additional sources, PS1 and PS2, 
found during the course of this analysis. 
The bright source to the east of the SMC is the known gamma-ray source associated with 47 Tuc (3FGL~J0023.9$-$7203). The map is binned in 0$\fdg$1$\times$0$\fdg$1
pixels smoothed with a $\sigma$ = 0$\fdg$3 Gaussian kernel. \label{fig:SMCcountsmap}}
\end{figure}

\begin{figure}[ht]
\includegraphics[width=0.6\columnwidth]{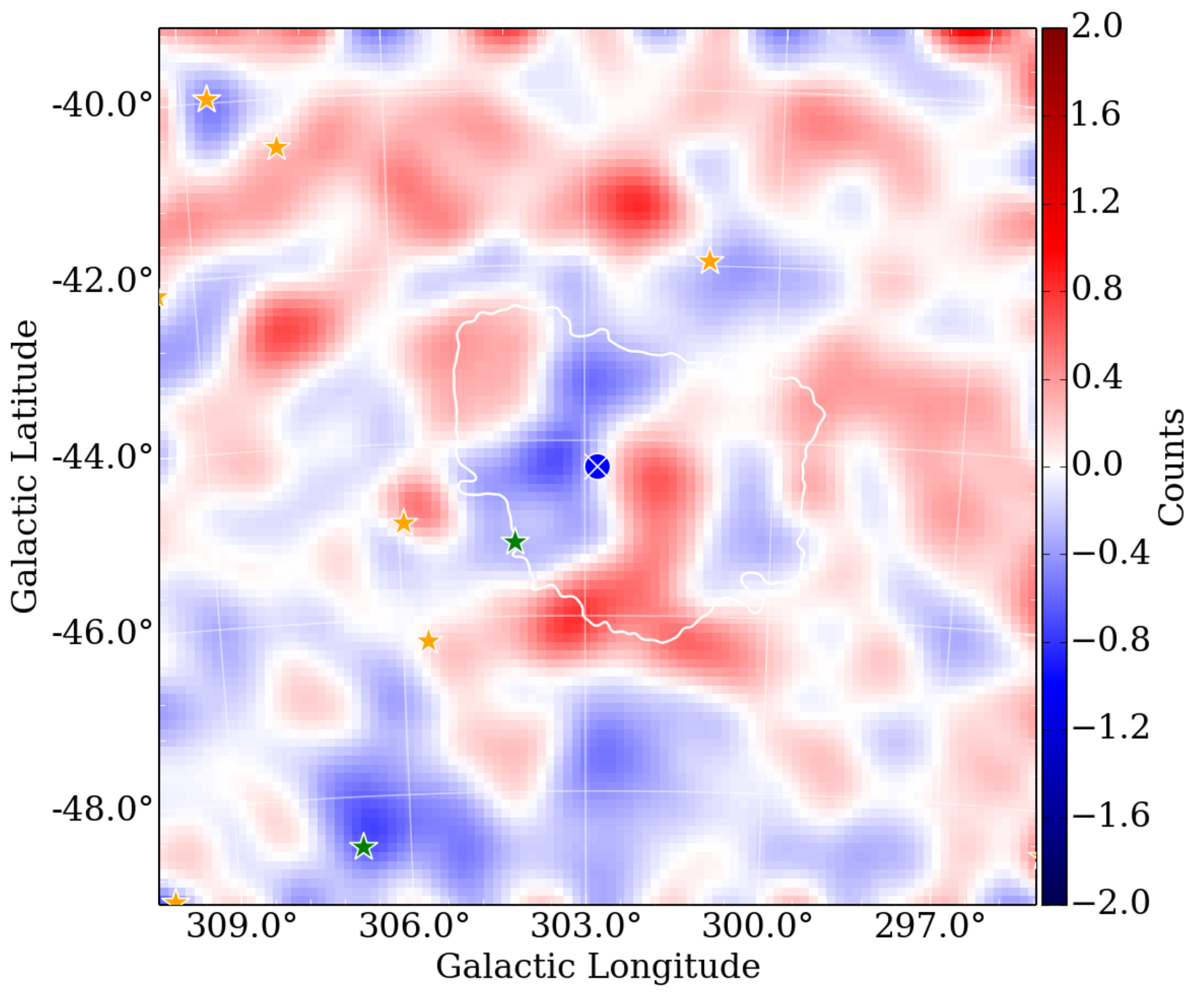}
\caption{Residual map (data - model) of gamma rays in the SMC region, for 0.5--500~GeV.
The diffuse emission of the SMC from Section~\ref{sec:smc} is outlined in the white solid contour representing 4\% of the maximum.
The kinematic center of the SMC is the blue circle with the white ``X" through it. 
Orange stars indicate sources in the 3FGL. Green stars indicate the two additional sources, PS1 and PS2,
found during the course of this analysis. The map is binned in 0$\fdg$1$\times$0$\fdg$1
pixels smoothed with a $\sigma$ = 0$\fdg$3 Gaussian kernel. \label{fig:SMCresmap}}
\end{figure}

The data reduction and exposure calculations were performed using the LAT \textit{ScienceTools} version 
10-01-01\footnote{\url{http://fermi.gsfc.nasa.gov/ssc/data/analysis/software}}. 
The event selection for the analysis is summarized in Table~\ref{tab:data}. 
In Figure~\ref{fig:SMCcountsmap}, we show a counts map of the gamma rays in the SMC ROI 
and in Figure~\ref{fig:SMCresmap} we show the residual (data - model) map, both for 0.5--500 GeV. 
The residual map is consistent with statistical fluctuations indicating the model is in agreement with the data.

\begin{table}[ht]
\begin{tabular}{cc}
\hline \hline
Selection & Criteria \\ \hline
Observation Period & 2008 Aug. 4 to  2014 Aug. 5\\
Mission Elapsed Time (s)\footnote{$Fermi$ Mission Elapsed Time is defined as seconds since 2001 January 1, 00:00:00 UTC} & 239557414 to 428903014 \\ 
Energy Range (GeV) & 0.5--500\\
Fit Region & 10$\dg\times$10$\dg$ centered on $(\ell,b)=(302\fdg80, -44\fdg30)$ \\ 
Zenith Range & $\theta_z<$100$\dg$\\
Data Quality Cut\footnote{Standard data quality selection: \texttt{DATA\_QUAL==1 $\&\&$ LAT\_CONFIG==1} with the $gtmktime$ Science Tool} &  yes \\ \hline
\end{tabular}
\caption{ \label{tab:data} Summary table of {\it Fermi}-LAT data selection criteria used for this paper's DM analysis. }
\end{table}

\section{ANALYSIS}\label{sec:analysis}

The analysis techniques we apply in this paper closely follow those used in the LMC analysis~\cite{Buckley:2015doa}. We take the center of the 
DM distribution ($\ell_{DM}$, $b_{DM}$) to be the kinematic center of the SMC (see Table~\ref{tab:data}). 
The parameters of the DM density profile, the final annihilation 
states and the mass ($m_{\chi}$) define the search parameters. If an excess is detected, our goal is to determine the DM
mass and annihilation cross section that best fits the observation. If no excess is found, we set an upper limit on the cross section ($\langle \sigma v \rangle$). 
The DM density profile which best fits both the SMC rotation curve and the results from the hydrodynamic cosmological 
simulation is the generalized NFW, or {\tt gNFW} (Section~\ref{sec:dm}).  
Here we also investigate the pure {\tt NFW} profile that best fits the SMC rotation curve. We first briefly 
describe the analysis technique we used to constrain the DM annihilation rate. Even though the SMC models fit the gamma-ray data, 
the models are empirically fit to the LAT data set. We must take into account degeneracies 
between our SMC modeling of the CR background and a potential DM signal. These systematic uncertainties are addressed later in the section.

\subsection{Fitting Method}\label{sec:fitting}
To fit the DM template we use a multi-step likelihood fitting procedure that has been previously applied to searches for 
DM signals in dwarf spheroidal galaxies~\cite{Ackermann:2015zua, Drlica-Wagner:2015xua}, the Smith High-Velocity 
Cloud~\cite{Drlica-Wagner:2014yca}, and the LMC~\cite{Buckley:2015doa}. 
We first perform a broadband fit over the entire energy range. 
This broadband fit determines the normalizations of the diffuse sources and of the 
point-like background sources within 10$\dg$ of the kinematic center of the SMC. 
We consider sources over a larger area than the ROI.
In this step, the spectral component of the DM annihilation gamma rays is modeled as a power law (index 
$\Gamma$ = 2)\footnote{Varying this power-law index by $\pm$0.5 does not significantly affect the results of the fit.}. The spectral shape of the DM in each 
annihilation channel is taken into account in the second step of the analysis. 

We then scan the likelihood as a function of the flux normalization of the assumed DM signal independently in each energy bin to create a spectral energy distribution 
for a source of the spatial morphology that we assume.  For this bin-by-bin scan, we fix the normalizations of the background sources to avoid instabilities 
resulting from fine binning in energy and correlations between the Galactic and isotropic diffuse components. 
By analyzing each energy bin separately, we avoid selecting a single spectral shape to span the entire energy range 
at the expense of introducing additional degrees of freedom into the fit. For the fit in any given bin, the only free parameter describing the DM component is the 
normalization. 

Since our background model is an empirical description of the SMC region, we must identify and quantify the degeneracies between 
the DM models and the components of the background model. We then allow the normalizations of those 
components to vary within the statistical uncertainties of the broadband fit when performing the bin-by-bin fitting. The degenerate components are determined 
based on their correlation with the spatial morphology of the DM template. This is described in Section~\ref{sec:correlations}. 

The uncertainties associated with each background component are derived from the results of the broadband fit. To account for the reduced statistics
in the bin-by-bin fits, we assign the width of the assumed Gaussian prior on the background components highly correlated with the dark-matter signal 
as ten times the uncertainties of the parameters in the 
broadband fit as determined empirically from a coverage study described in Section~\ref{sec:correlations} 
\footnote{We varied the width of the Gaussian prior and found that using a factor of ten resulted in the correct coverage properties
for upper limits on the Monte Carlo.}. By allowing the background normalizations in 
the bin-by-bin fit to vary within uncertainties, we can estimate not only the correlations between the 
background and signal models but also the significance of any observed excess. From the significance, we can also calculate the upper limit on the cross section for a 
specific final state as a function of DM mass. 

We evaluate the significance of the DM hypothesis using the test statistic (TS) defined as: 
\begin{equation}
\mathrm{TS}=2~\mathrm{ln}\frac{\mathcal{L}(\mu, \theta | \mathcal{D})}{\mathcal{L}_{\mathrm{null}}(\theta | \mathcal{D})} \label{eq:ts}
\end{equation}

\noindent For DM masses up to $\sim$500 GeV the statistics are large enough that the TS-distribution follows a $\chi^{2}$ distribution (Chernoff's theorem~\cite{REF:Chernoff}) 
and the significance of a given TS value can be calculated from the tail probability of the $\chi^{2}$ distribution function. 
As the counts per bin decreases, the $\chi^{2}$ distribution moderately over-predicts the number of high TS trials observed 
in simulated data. 

The final step of the fitting procedure is to convert the bin-by-bin likelihood curve in flux into a likelihood curve in $\langle \sigma v \rangle$ for each 
spatial profile and annihilation channel, which determines the spectrum. We scan DM masses ($m_{\chi}$) from 2--10000 GeV (when kinematically allowed in the annihilation channel under consideration), and 
the pair-annihilation final states listed in Section~\ref{sec:dm}. For each DM spectrum, we extract the expected flux, $F_{j}$, in each energy bin and calculate
the likelihood of observing that flux value. The log-likelihood in each energy bin is summed to get the log-likelihood curve, defined as: 

\begin{equation}
\mathrm{ln}~\mathcal{L}(\langle \sigma v \rangle, \mu, \theta | \mathcal{D}) = \sum_{j} \mathrm{ln}~\mathcal{L}_{j}(\langle \sigma v \rangle, \mu, \theta_{j} | \mathcal{D}_{j})  \label{eq:llcurve}
\end{equation}

\noindent For each DM mass and channel, we can calculate the maximum likelihood cross section and the 95\% CL upper limit on the cross section. 

\subsection{Correlations Between Background and Signal Components}\label{sec:correlations}

\begin{figure}[ht]
\includegraphics[width=0.48\columnwidth]{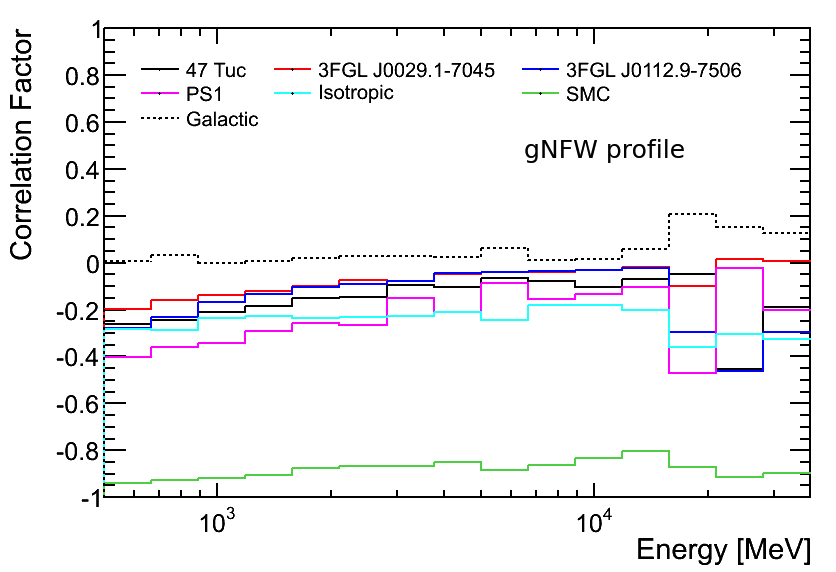}
\includegraphics[width=0.48\columnwidth]{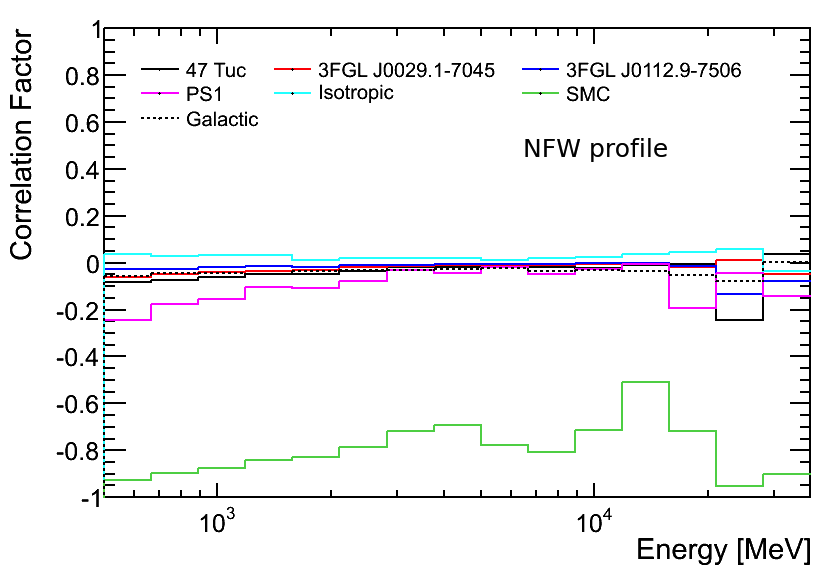}
\caption{Correlation factors, as given by Eq.~\ref{eq:correlation}, between the most important components of the background model and the two 
DM density profiles, {\tt gNFW} (left) and {\tt NFW} (right), as a function of photon energy. \label{fig:correlation}}
\end{figure}

To measure the effect of correlations between background models and the simulated DM signal, in our bin-by-bin fit we allowed 
the normalizations of different backgrounds to vary within ten times the uncertainties of the parameters in the broadband fit. 
The background components we tested for correlations were those nearest the kinematic center of the SMC and are as follows: 
the new point source (PS1), two 3FGL sources (3FGL J0029.1$-$7045 and 3FGL J0112.9$-$7506), 47 Tuc, 
the extended SMC component, and the isotropic and the Galactic diffuse backgrounds. The correlation factor at a given energy between the DM
component and the $i^{th}$ background component in energy bin $j$ can be obtained from the covariance matrices for the parameters once the likelihood function has been maximized: 

\begin{equation}
\rho_{i,\mathrm{DM}}(j) = \frac{\mathrm{cov}_{i,\mathrm{DM}(j)}}{\sigma_i(j) \sigma_{\mathrm{DM}}(j)}  \label{eq:correlation}
\end{equation}

\noindent where $\sigma_i(j)=\sqrt{\mathrm{cov}_{i,i}(j)}$ is the variance on the normalizations of the $i^{th}$ model component in the $j^{th}$ energy bin. 
In Figure~\ref{fig:correlation}, we show the correlation factor between the dark matter and background component as a function of photon energy 
for the {\tt gNFW} and {\tt NFW} DM profiles at the kinematic center of the SMC. We found that the SMC component, PS1, 
and the isotropic diffuse have the highest correlation factor with the {\tt gNFW} DM profile, whereas the {\tt NFW} 
is most highly correlated with the SMC component only. Since PS1 has a relatively small flux, also allowing the normalization 
to vary in the bin-by-bin fit does not significantly change the results. 

\begin{figure}[ht]
\includegraphics[width=0.48\columnwidth]{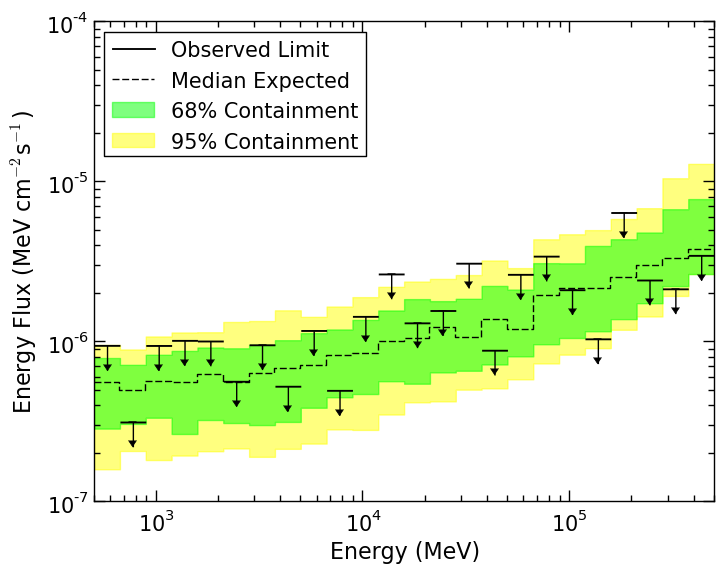}
\includegraphics[width=0.48\columnwidth]{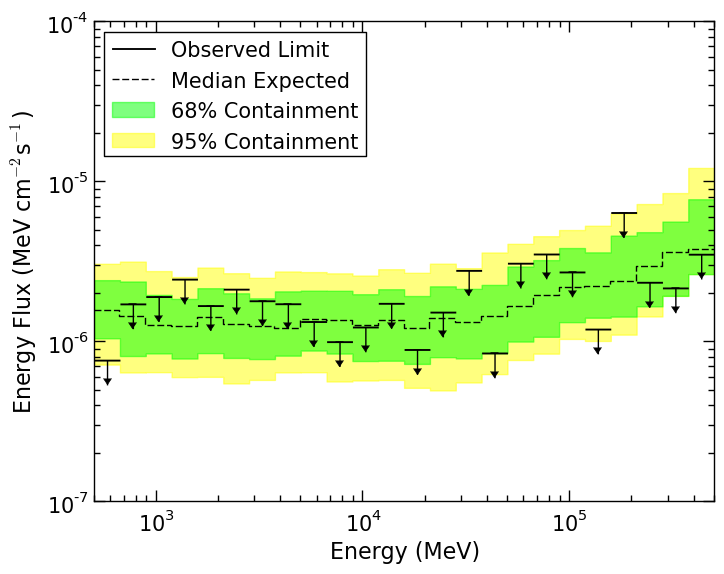}
\caption{95\% CL upper limit of the gamma-ray flux associated with a {\tt gNFW} density profile at the kinematic center of the SMC. 
In the left-hand panel, we use the values from the broadband fit to define the normalizations for the SMC component and the isotropic diffuse component. 
In the right-hand panel, we allow these normalizations to vary within the uncertainties (see text). The upper limits 
were determined using an MC simulation of the SMC ROI drawn with no DM contribution. The expected 1$\sigma$ (green) and 2$\sigma$
(yellow) containment bands for the upper limits are also shown. \label{fig:MCsed}}
\end{figure}

For the components that are highly correlated with DM, fixing the normalizations in the bin-by-bin analysis to the values derived from the broadband fit could 
result in a potential DM signal being assigned to one of the known backgrounds. This would result in an overly optimistic set of bounds on DM 
annihilations since there are large correlations with several of the SMC background components. 
In Figure~\ref{fig:MCsed}, we show the simulated energy bin-by-bin 95\% CL exclusion limits for an energy flux from a DM signal with a {\tt gNFW} 
morphology. In the left-hand panel, we use the values from the broadband fit to define the normalizations for the SMC component and the isotropic diffuse component. 
In the right-hand panel, we allow these normalizations to vary within the uncertainties. Especially at lower energies, 
the flux upper limits are significantly weaker when the normalizations are allowed to vary. 
This is due to the uncertainty associated with the correlation between the morphologies. 
To evaluate the upper limits, we generated a Monte Carlo (MC) simulation of the SMC ROI drawn from the background-only model outlined in Section~\ref{sec:dm} 
and the fitting procedure outlined above. We used 100 trials to construct the expected containment bands for the upper limits. 

\begin{figure}[ht]
\centering
  \includegraphics[width=0.48\columnwidth]{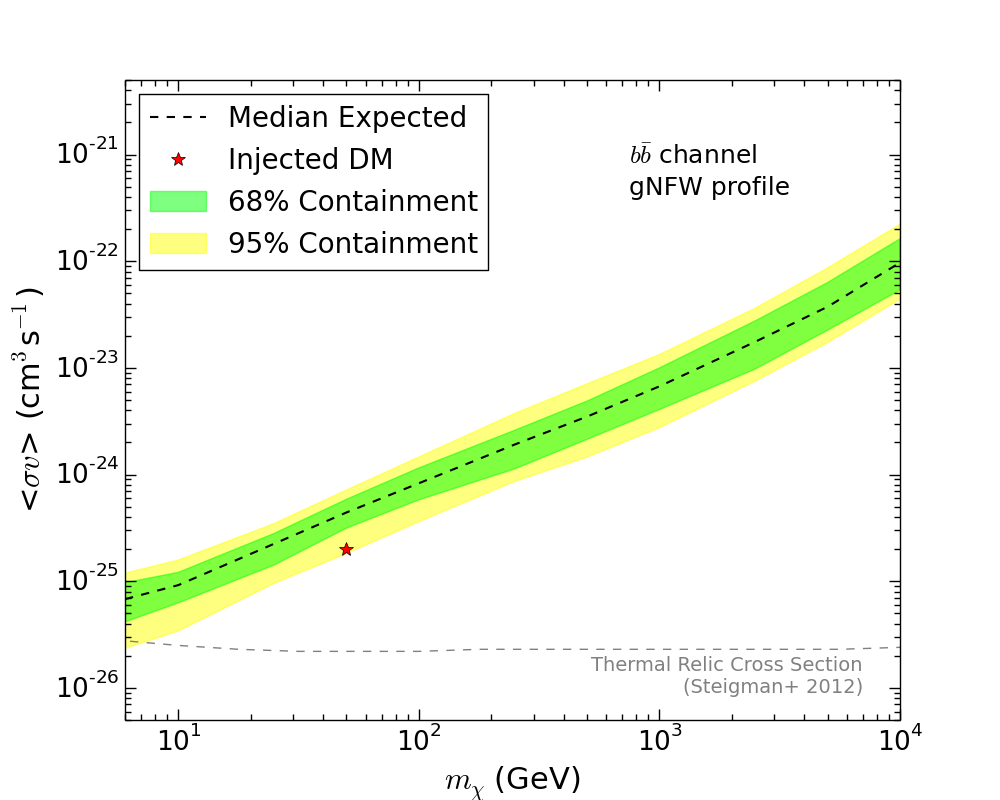}
  \includegraphics[width=0.48\columnwidth]{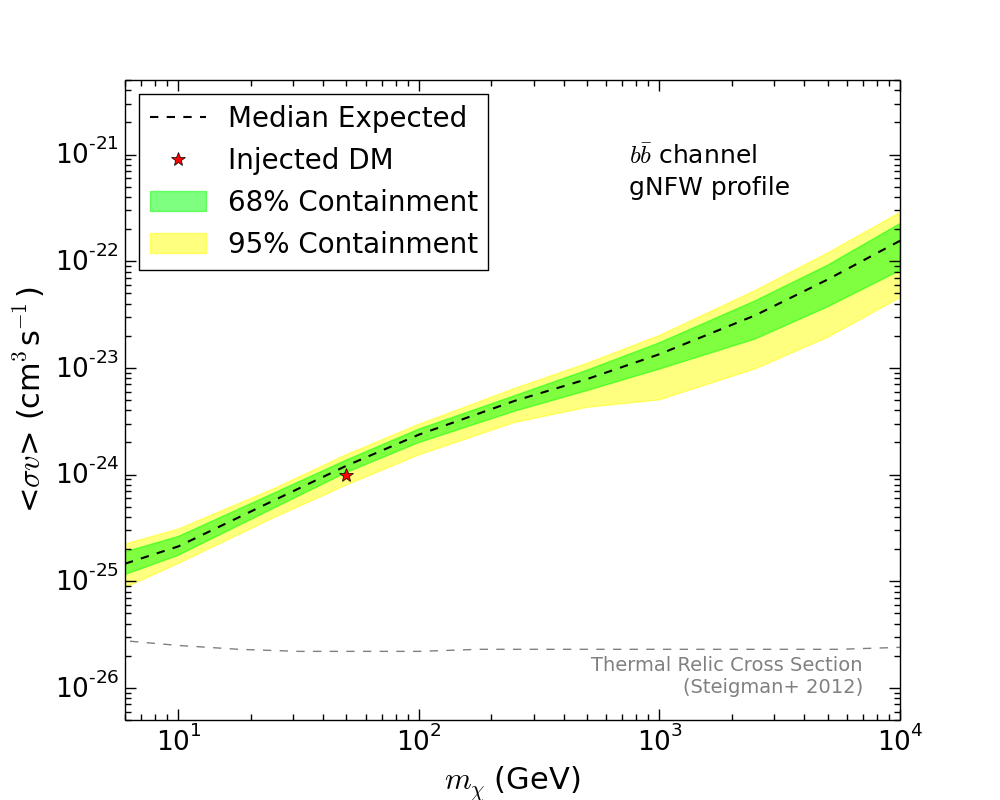}
\caption{Upper limit expectation bands in the presence of an injected DM signal. 68\% (green) and 95\% (yellow) containment bands for the 95\% CL upper limit on the 
annihilation cross section ($\langle \sigma v \rangle$) as a function of DM mass in the $b\bar{b}$ channel for the {\tt gNFW} profile centered at the kinematic center of the SMC. 
The bands are evaluated from 100 trials of MC simulations of the SMC background models with an injected signal of 50 GeV DM
annihilating into $b\bar{b}$ with a cross section of $\langle \sigma v \rangle$ = 2$\times$10$^{-25}$ cm$^{3}$/s (left) and 
$\langle \sigma v \rangle$ = 1$\times$10$^{-24}$ cm$^{3}$/s (right), shown as a star in the figures. The plots illustrate that 
the injected DM is not excluded. The horizontal dashed line shows the canonical thermal relic cross section~\cite{Steigman:2012nb}.}
\label{fig:coveragestudy}
\end{figure}

We calculate the expected exclusion curves for a DM annihilation signal from the kinematic center of the SMC by including a {\tt gNFW} profile 
with the simulation of the other components to demonstrate the coverage of our upper limit calculations. 
The injected signal\footnote{We tested 50 GeV DM annihilating to $b\bar{b}$ and $\tau^+\tau^-$ with 
$\langle \sigma v \rangle$ = 2$\times$10$^{-25}$ cm$^{3}$/s and $\langle \sigma v \rangle$ = 1$\times$10$^{-24}$ cm$^{3}$/s. 
We also tested 5 GeV DM annihilating to $b\bar{b}$ with $\langle \sigma v \rangle$ = 2$\times$10$^{-25}$ cm$^{3}$/s.} 
should lie below the 95\% CL upper limit on the cross section in 95\% of the pseudo-experiments. 
This is demonstrated in Figure~\ref{fig:coveragestudy}, where the injected signal falls at nearly the 95\% CL upper limit 
in the case with a smaller $\langle \sigma v \rangle$. The results were similar for the other injected signals.
When a larger DM annihilation signal is injected, the exclusion rate is higher (between the 68\% and 95\% CL upper limit bands). 
The consistency between the injected signal and the observed upper limits demonstrates that our method for setting upper limits has the correct frequentist coverage. (Note:
$\langle \sigma v \rangle$ = 1$\times$10$^{-24}$ cm$^{3}$/s is 50 times the nominal thermal relic cross section and has already been well excluded by other 
searches~\cite{Ackermann:2015zua}).

As an additional way to quantify the correlation between the DM annihilation term of the model and the background components of the model, 
we have adopted a technique similar to the LMC analysis 
to estimate the ``effective background"~\cite{Buckley:2015doa} (\textit{i.e.\@}, the background that overlaps with the signal~\cite{2014JCAP...10..023A}).
We calculate the ratio of signal events to effective backgrounds and compare it to the statistical uncertainty.
Since the only free parameters in the fit are the normalizations, the statistical uncertainty on the signal is 
$\sigma_{\mathrm{stat, sig}} \simeq \sqrt{b_{\mathrm{eff}}}$.

If the signal and background models are degenerate, the $b_{\mathrm{eff}}$ diverges
indicating that the model has little power to distinguish signal from background.
Comparing the expected statistical ($\sqrt{b_{\mathrm{eff}}}$) and systematic uncertainties ($\sim$0.02$\times b_{\mathrm{eff}}$ using 
Eq. 16 from~\cite{Buckley:2015doa}) 
we can determine if the analysis is statistics or systematics limited. 
For each source component, the effective background and expected counts ($N$) are shown in Table~\ref{tab:beff}, 
along with the total effective background counts for the model with the gNFW profile.

\begin{table}[ht]
\begin{tabular}{lcc}
\hline \hline
Source 				&  $b_{\mathrm{eff}}$ 	&$N$		\\ \hline
Isotropic diffuse 		& 5100 				& 8600 	\\
Galactic diffuse 		& 6400 				&18000 	\\ 
SMC  				& 140 				& 2000 	\\
PS1	 				& 81.7 				& 24.8	\\
47 Tuc 				& 0.0043 				& 2150	\\ 
PS2					& 0.02 				& 870	\\ \hline
Total 	 			& 25300  				& 31600	\\
\end{tabular}
\caption{The results of the $b_{\mathrm{eff}}$ calculation for each component of the background model with the total 
$b_{\mathrm{eff}}$ and the number of expected counts ($N$).\label{tab:beff}}
\end{table}

\subsection{Alternative SMC Background Modeling}\label{sec:sys}

\begin{figure}[ht]
\centering
  \includegraphics[width=0.6\columnwidth]{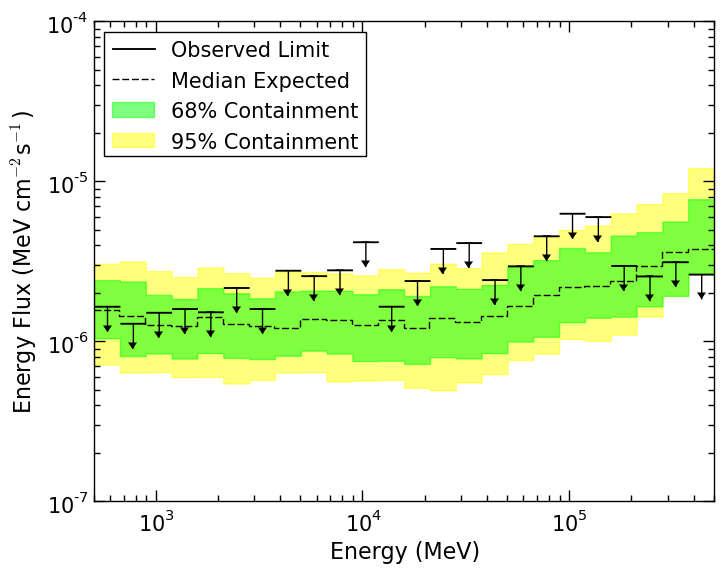}
\caption{The flux upper limits of the ``alternative" (2D Gaussian) background model for the SMC.  
The bands are derived from the emissivity model of the SMC. }
\label{fig:2DSMCmodel}
\end{figure}

In Section~\ref{sec:smc}, we described two methods for modeling the SMC. For the main results of this analysis we used 
the model that provided the overall maximum likelihood, the one-component emissivity model. We repeated the analysis using 
the alternative 2D Gaussian to model the SMC. 
This allows the sensitivity of the DM limits to the background model of the SMC to be estimated. 
We allowed the isotropic diffuse and the 2D Gaussian model to vary within ten times the statistical uncertainties as determined
by the broadband fit when we introduced a DM component. In Figure~\ref{fig:2DSMCmodel}, we show the resulting flux upper limits.
For the bands we used the same MC simulation derived from the emissivity model shown in the right-hand panel of Figure~\ref{fig:MCsed}. 
The flux upper limits of the 2D Gaussian SMC model are mostly within the two sigma bands of the MC simulation of the emissivity SMC model 
illustrating sufficient agreement between the two models. 

We find that overall the largest contributor to the systematic uncertainty is the correlations with the backgrounds as described in Section~\ref{sec:correlations}.  
These uncertainties are large enough that in our final results we do not explicitly consider the modeling uncertainties suggested by the comparison we made here. 

\section{RESULTS}\label{sec:results}

We can now set constraints on the annihilation of DM into Standard Model particles
that result in gamma rays. We report the 95\% CL upper limit on the annihilation cross section for the 
$b\bar{b}$, $\tau^+\tau^-$ channels. These channels were previously considered in the dwarf spheroidal 
analyses~\cite{Ackermann:2015zua, Drlica-Wagner:2015xua}. 
We evaluate cross section limits for the {\tt gNFW} and the {\tt NFW} DM distributions considered in Section~\ref{sec:dm}.
Since these two distributions also represented the highest and lowest $J$-factor estimates, the results can also be interpreted
as optimistic and conservative limits respectively. 

\begin{figure}[ht]
\includegraphics[width=0.48\columnwidth]{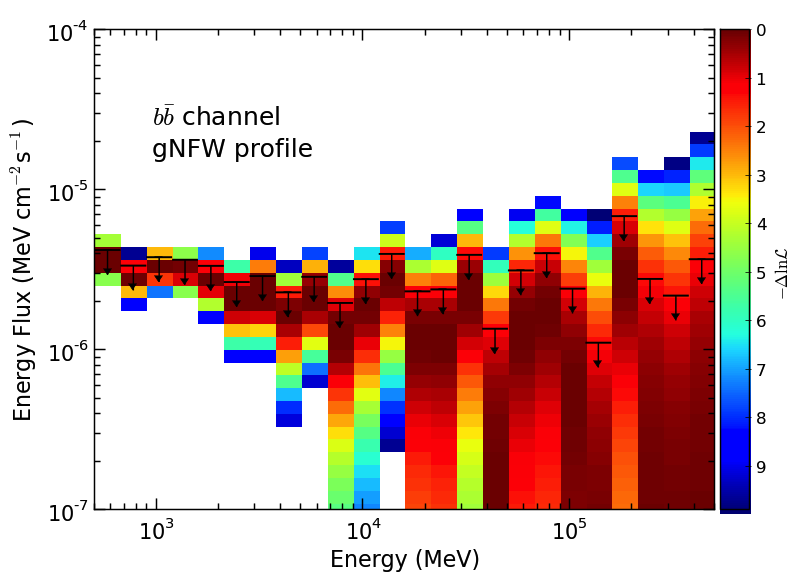}
\includegraphics[width=0.435\columnwidth]{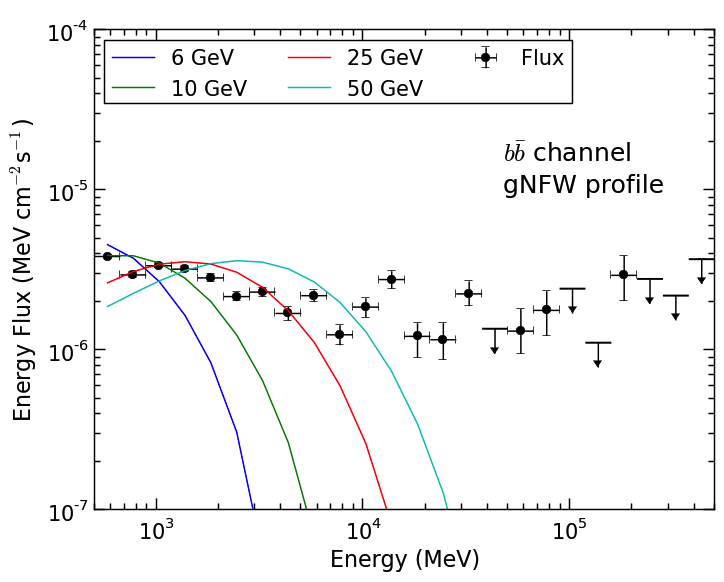}
\caption{The left-hand panel shows the change in the logarithm of the likelihood as a function of the flux of the DM component 
which follows a gNFW profile centered at the SMC kinematic center. 
The profiles are truncated at $-\Delta$ln$\mathcal{L}>$10. Upper limits on the integrated energy flux are set at 95\% CL
within each bin where the log-likelihood has decreased by 2.71/2 from its maximum~\cite{Ackermann:2013yva}.
The right-hand panel shows the Maximum Likelihood Estimator (MLE) of the energy flux per energy bin. At lower energies, the flux is 
fit with four different DM masses in the $b\bar{b}$ channel (similar results were obtained for the $\tau^+\tau^-$ 
channel). The resulting best fits of the normalizations correspond to a $\langle \sigma v \rangle$
for each mass; however, the $\langle \sigma v \rangle$ for all four DM masses and both channels have already been excluded~\cite{Ackermann:2015zua}. 
\label{fig:noSMCgNFWlimits}}
\end{figure}

Since the one component emissivity background model of the SMC is highly correlated with both the {\tt gNFW} and {\tt NFW} profiles (see Figure~\ref{fig:correlation})
and the model is data driven, we first wanted to measure $\langle \sigma v \rangle$ under the assumption that all the gamma rays from the SMC
region are attributable to DM annihilation. 
We evaluated the flux dependence of the maximum of the likelihood function, shown in the left-hand panel of Figure~\ref{fig:noSMCgNFWlimits}.
As expected, when only a {\tt gNFW} template is used (and the SMC template is neglected), 
the likelihood analysis would indicate a significant DM component, especially in the low energy ($<$10 GeV) bins.
If this excess could be attributed entirely to DM annihilation it would follow the predicted spectral energy distribution for that process. 
The right-hand panel shows the Maximum Likelihood Estimator (MLE) of the energy flux per energy bin. The MLE prediction of the flux is 
fit to a DM spectral template where the normalization ($\langle \sigma v \rangle$) is fit with four different DM masses. 
The fit is performed by assembling a global likelihood from the bin-by-bin likelihoods and the global likelihood is maximized to derive the normalization. 
In this figure, only the $b\bar{b}$ channel is shown; however, similar results were obtained for the $\tau^+\tau^-$ channel. 
Since a pure DM annihilation spectrum is not a good fit to the observed spectrum, the majority of the gamma rays from the SMC 
are not from DM annihilation. Furthermore, the implied MLE cross 
sections are large and have been excluded by the dwarf spheroidal DM searches~\cite{Ackermann:2015zua}. This 
further supports the conclusion that the gamma-ray emission from the SMC is not dominantly from DM annihilation unless 
the $J$-factor is implausibly larger than the values we derived.

\begin{figure}[ht]
\includegraphics[width=0.48\columnwidth]{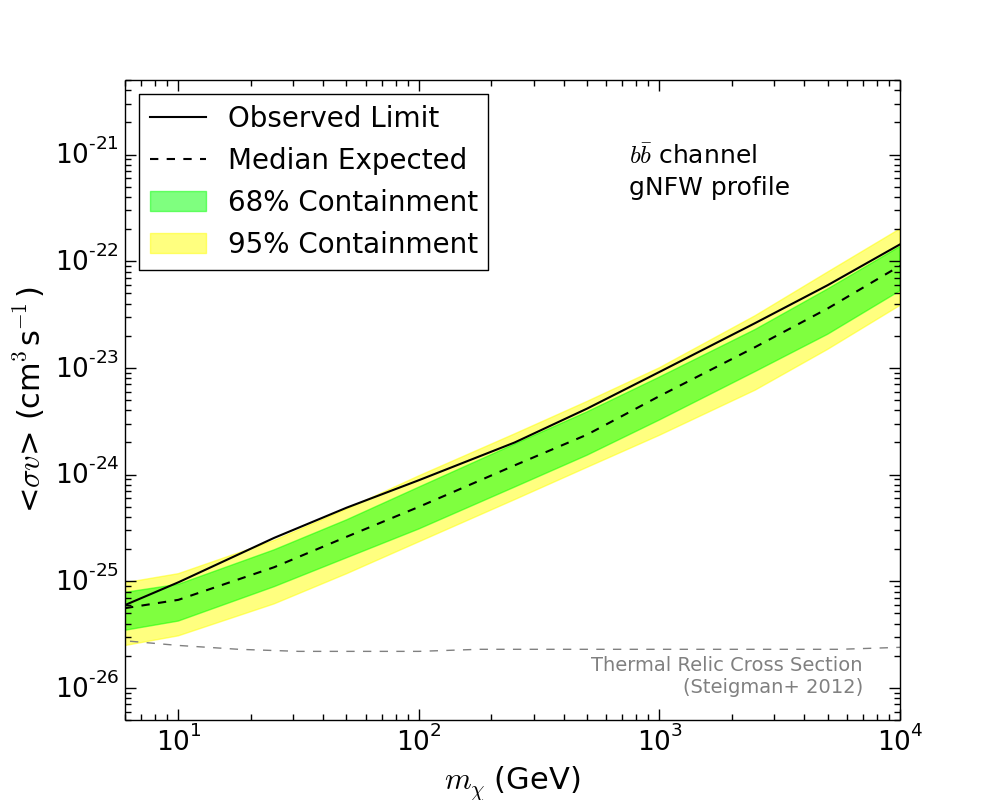}
\includegraphics[width=0.48\columnwidth]{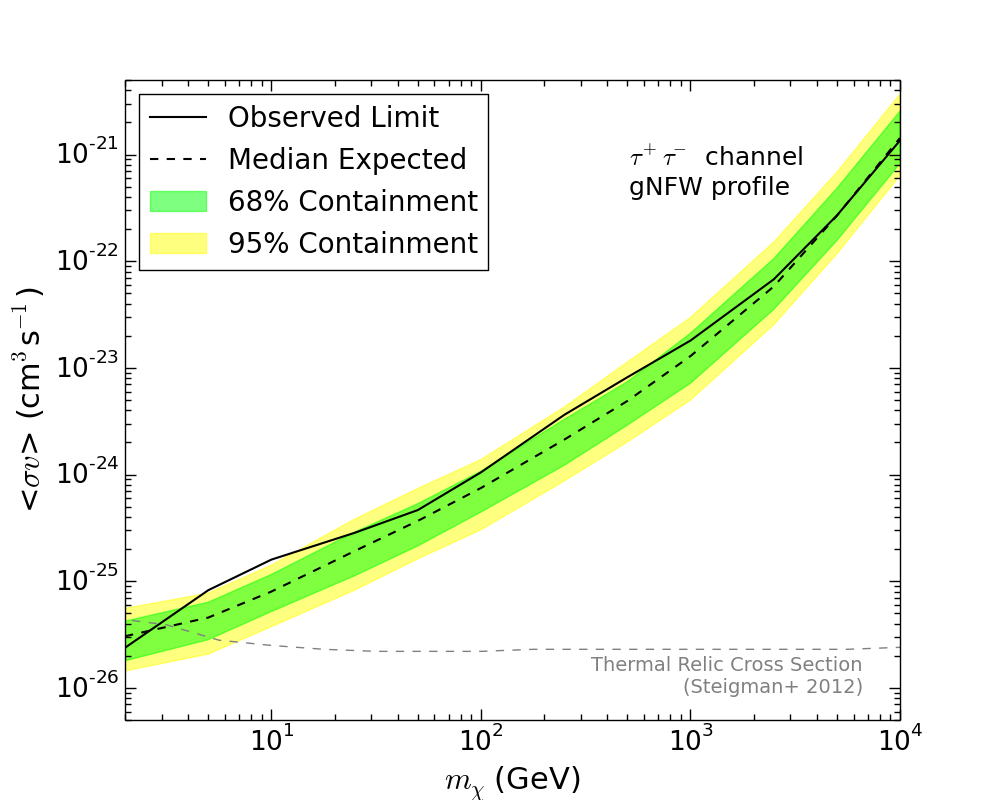}
\caption{Upper limits for $\langle \sigma v \rangle$ for the $b\bar{b}$ and $\tau^+\tau^-$ annihilation channels 
(solid black), as a function of DM mass, assuming the {\tt gNFW} profile located at the SMC kinematic center. Also shown are the 68\% (green) and 95\% (yellow) containment bands of the upper
limits drawn from background-only simulations. The horizontal dashed line shows the canonical thermal relic cross section~\cite{Steigman:2012nb}.  \label{fig:gNFWlimits}}
\end{figure}

\begin{figure}[ht]
\includegraphics[width=0.48\columnwidth]{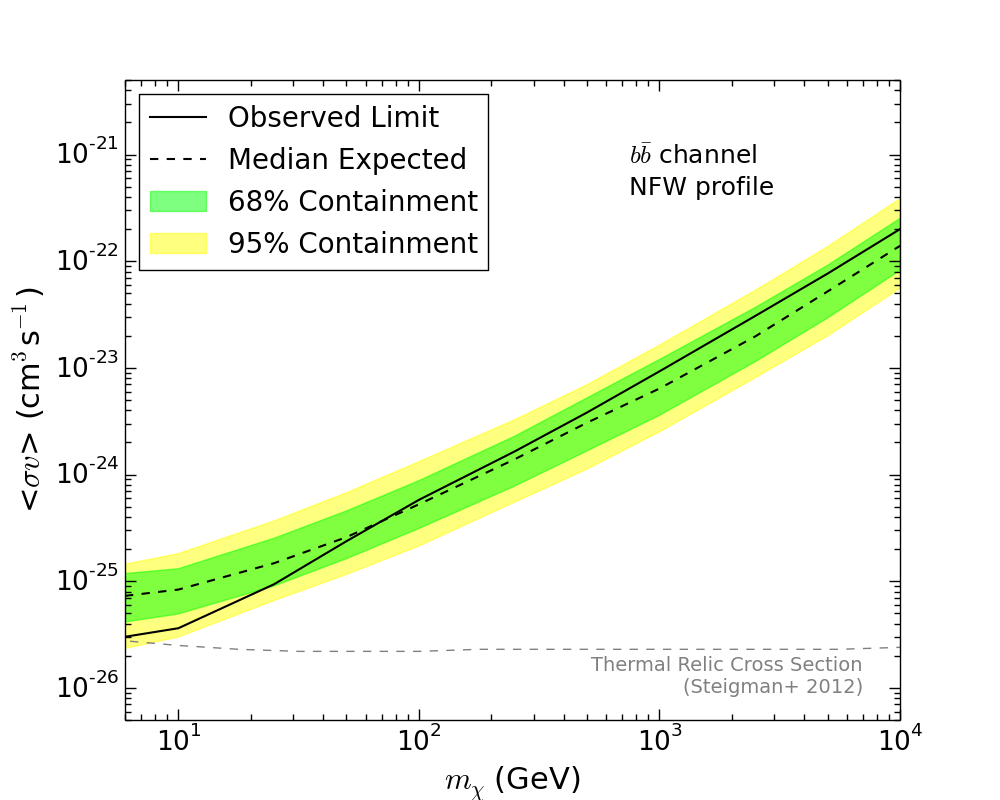}
\includegraphics[width=0.48\columnwidth]{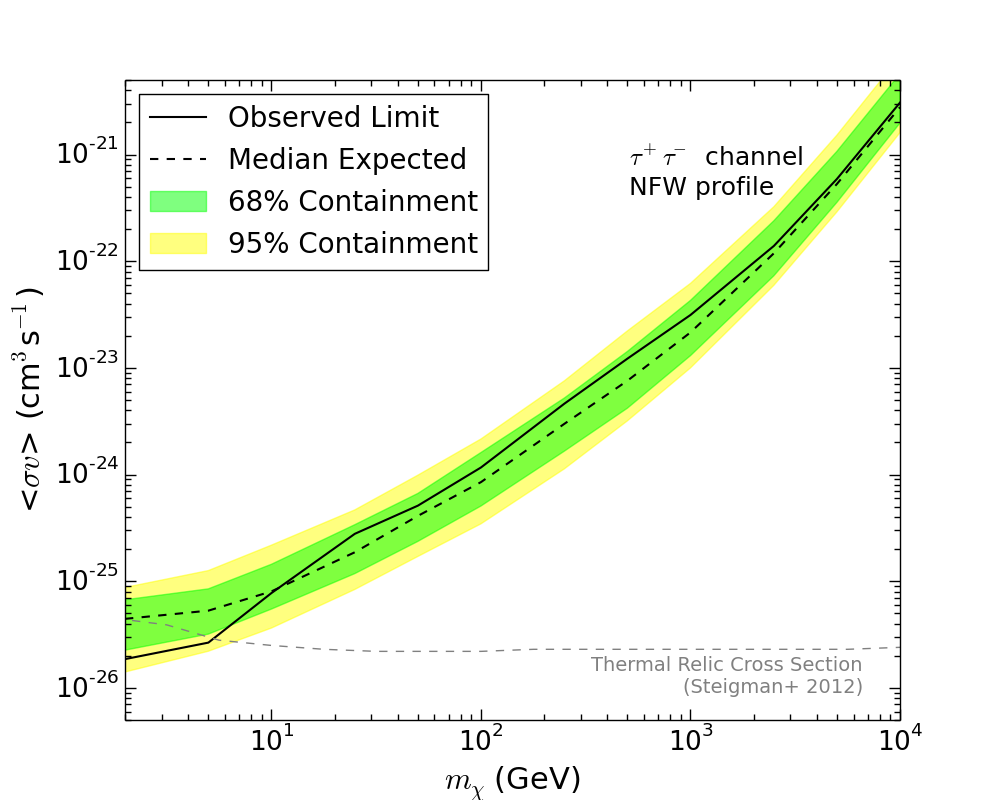}
\caption{Upper limits for $\langle \sigma v \rangle$ for the $b\bar{b}$ and $\tau^+\tau^-$ annihilation channels 
(solid black), as a function of DM mass, assuming the {\tt NFW} profile located at the SMC kinematic center. Also shown are the 68\% (green) and 95\% (yellow) containment bands of the upper
limits drawn from background-only simulations. The horizontal dashed line shows the canonical thermal relic cross 
section~\cite{Steigman:2012nb}. \label{fig:NFWlimits}}
\end{figure}

Since a pure DM annihilation spectrum is not a good fit to the observed spectrum of the SMC region and the implied DM cross section has already been excluded by other studies, 
we include the emissivity model of the SMC to derive upper limits on the annihilation cross sections for the two choices for the DM density profile. 
In Figures~\ref{fig:gNFWlimits} and~\ref{fig:NFWlimits} we show the 95\% CL upper bounds for the 
$b\bar{b}$ and $\tau^+\tau^-$ annihilation channels for the {\tt gNFW} and {\tt NFW} profiles. 
There are no obvious features in the mass dependence of these limits, except the dip in the observed limit at lower DM
masses for the {\tt gNFW} profile. This could result from a fluctuation in the data, or an overestimation of 
the correlation of the DM signal with the backgrounds at low masses. The highest TS value measured is $\sim$4 in the $\tau^+\tau^-$ channel using the 
{\tt NFW} profile. This is shown as the slight excess in Figure~\ref{fig:gNFWlimits} at 10 GeV.
In these figures, we also see that the two profiles set nearly the same constraints in both channels. 
This is not surprising since, although the {\tt gNFW} profile has a larger $J$-factor than the {\tt NFW} profile, it is also more highly correlated 
with the SMC baryonic template (especially at higher masses). In the case of 
the {\tt NFW} profile, we can exclude the canonical thermal cross section for DM up to 
$\sim$7 GeV in the $\tau^+\tau^-$ channel. This is consistent
with the results from the dwarf spheroidal and LMC analyses~\cite{Ackermann:2015zua, Drlica-Wagner:2015xua, Buckley:2015doa}. 
When using the {\tt gNFW} profile, we cannot exclude the canonical thermal cross section in the $b\bar{b}$ channel at any mass, 
and only up to $\sim$2 GeV in the $\tau^+\tau^-$ channel. 
Several studies have shown that in the Galactic center when the annihilation channel has a large leptonic component 
(i.e., the $\tau^+\tau^-$ case), secondary emission is important to consider in particular at high masses ($m_{\chi}>$1000 GeV)~\cite{Buch:2015iya, Kaplinghat:2015gha, Lacroix:2015wfx, 
Gomez-Vargas:2013bea, Ackermann:2015tah}. 
In the case of the SMC, the strengths of both the magnetic field and the Interstellar Radiation Field (ISRF) are smaller than in the 
Galactic center~\cite{2010A&A}, and the gas densities are not any greater.  
It is most likely that the signal from the secondaries would have a similar spatial distribution as the other classical backgrounds, 
making it difficult to disentangle the signal and background components in the fit.
We estimate that if the secondary emission was distinguishable from classical backgrounds, considering it would improve the limits at 
$E_{\gamma}<$1 GeV and for $m_{\chi}>$1000 GeV by 20\%, which is subdominant to other statistical and systematic effects.

The expectation bands in Figures~\ref{fig:gNFWlimits} and~\ref{fig:NFWlimits} are derived from the same 100 MC trials that were used in deriving the flux upper limits 
in the right-hand panel Figure~\ref{fig:MCsed}. See Section~\ref{sec:correlations}.
This correlation is the largest uncertainty in our limits, in particular at energies 
less than 100 GeV. We find that the observed limits are weaker than the predicted limits over most of the mass range for 
both channels when using the {\tt gNFW} spectral model for the DM component. 
This indicates that letting the background components most correlated with the DM signal vary within uncertainties in the bin-by-bin fit is 
not a perfect method to take the correlation of the components into account. However, our observed limits using the 
{\tt NFW} profile are consistent and in some cases more constraining than indicated by the MC simulations. 
  
\section{DISCUSSION AND CONCLUSIONS}\label{sec:conclusion}

The SMC is the second largest satellite of the Milky Way and contains an amount of DM
that would result in a DM annihilation signal at Earth as large as the brightest dwarf galaxies. Although it has less DM 
than the LMC, and is slightly more distant, it also has less interstellar gas and massive star formation, and fewer conventional sources of gamma rays. 

Given the enormous interest in the possible detection of a DM signal from 
the Galactic center, the advantages of the SMC makes it an excellent target for analysis with $Fermi$-LAT gamma-ray 
data. This work provides the first constraints on the annihilation of DM into Standard Model particles
from observations of the SMC. 
Consideration of both hydrodynamical cosmological simulations and the observed rotation curve indicates that the DM density 
profile is best described by either a {\tt gNFW} or an {\tt NFW} profile. We based our search for a DM annihilation signal on six years of $Fermi$-LAT 
data over a 10$\dg\times$10$\dg$ ROI centered at the SMC. To derive a model for the gamma-ray emission from the SMC, 
we employed a physical emissivity model which yielded a one-component description of the SMC ({\it Fermi}-LAT Collaboration, 2016, in prep.). 
We place upper limits on the velocity-averaged cross section that reach the benchmark canonical thermal freeze-out value for DM with an {\tt NFW} up to 
$\sim$7 GeV in the $\tau^+\tau^-$ channel. 
Compared to the expectation given the SMC baryonic model and the DM profiles, the limits we found were modestly weaker than expected. 
No DM annihilation signal was found to be statistically significant, the largest being slightly greater than 2$\sigma$. 
An interpretation of the significance should also consider a trials factor, which further reduces the value.

The main source of uncertainty is correlations between the SMC gamma-ray emission model and the {\tt gNFW} and {\tt NFW} DM profiles. 
We found that these DM profiles are highly correlated with components of our SMC background model and this is the largest source of uncertainty. 
The correlation between these components weakens the limits, 
in particular in the energy range associated with the excess near the Galactic center.

\begin{figure}[ht!]
\centering
  \includegraphics[width=0.6\columnwidth]{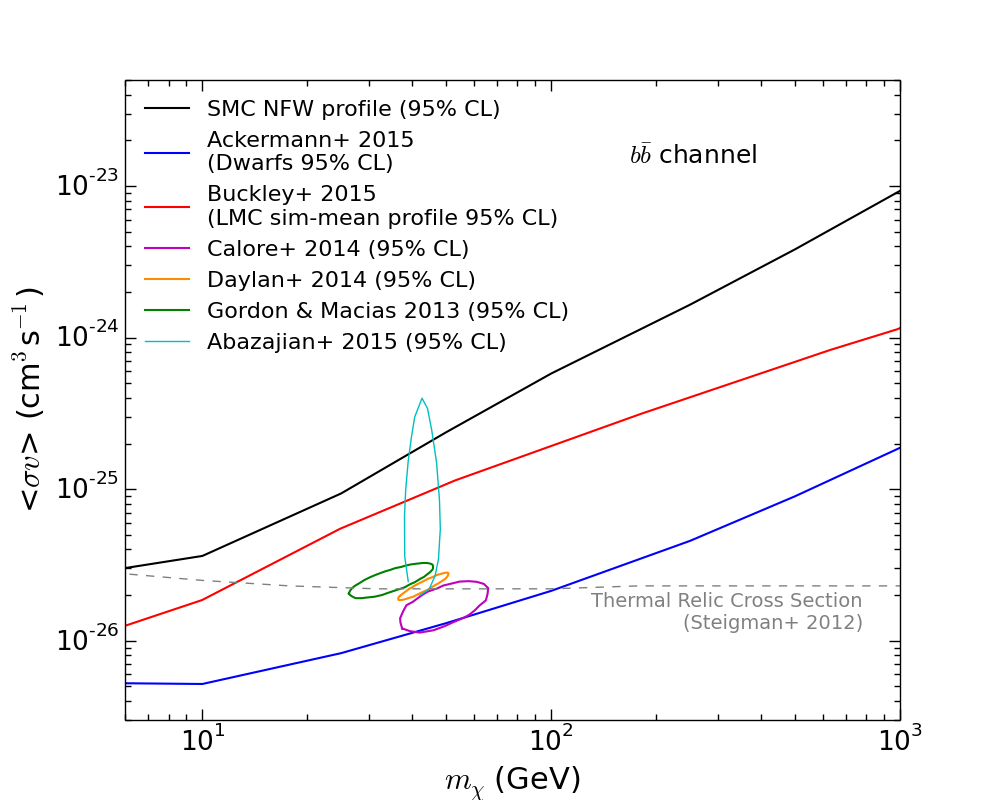}
\caption{A comparison between the 95\% CL upper limits from the SMC analysis (black solid line), the LMC analysis~\cite{Buckley:2015doa}, 
and the $Fermi$-LAT analysis of the dwarf spheroidal galaxies~\cite{Ackermann:2015zua}. Also shown are the confidence regions 
for cross section and mass determined by analyses of the Galactic center excess. These are shown in solid magenta~\cite{Calore:2014zzz}, 
orange~\cite{Daylan:2014rsa}, green~\cite{Gordon:2013vta}, and cyan~\cite{Abazajian:2015raa} respectively. The SMC upper limits are based on the 
{\tt NFW} profile. The results are similar with both profiles used in the analysis. 
The horizontal dashed line shows the thermal relic cross section~\cite{Steigman:2012nb}.}
\label{fig:Combined}
\end{figure}

Both DM profiles yielded similar limits and are either competitive with or exceed the existing limits from any individual 
dwarf spheroidal galaxy. However, the limits are weaker than the limit from a joint analysis of the dwarf spheroidal galaxies. 
The limits we found are also comparable to the more conservative limits derived from the analysis of the LMC. 
In Figure~\ref{fig:Combined}, we compare the bounds set for the $b\bar{b}$ channel using the {\tt NFW} profile, 
the LMC analysis which used {\tt Pass 7 Rep} data and a more realistic {\tt sim-mean} profile~\cite{Buckley:2015doa}, 
and the most recent $Fermi$-LAT analysis of the dwarf spheroidal galaxies~\cite{Ackermann:2015zua}. We expected to find
stronger or comparable bounds across the entire mass range, but did not due to correlations of the DM profiles with components of the SMC background.
In the figure, we also show our limits compared with the values preferred by analyses of the Galactic center excess~\cite{Calore:2014zzz, 
Daylan:2014rsa, Gordon:2013vta, Abazajian:2015raa}. 
The ellipses shown in the figure are meant to illustrate the parameter space of interest and do not include
uncertainties on  $\langle \sigma v \rangle$ due to the uncertainties of the corresponding DM density profiles.

A better understanding of the populations of cosmic rays and high energy sources in the SMC would help disambiguate 
astrophysical emission from DM signals.
In this study, our approach was conservative, resulting in robust bounds. More accurate simulations of the Magellanic system eventually will be made possible 
using the results of stellar surveys, such as GAIA~\cite{Perryman:2001sp}, and would give greater confidence in both the 
morphological shape of the expected DM signal and value of the $J$-factor. Nevertheless, 
the present work and the LMC analysis~\cite{Buckley:2015doa} already demonstrate the potential of DM searches in complicated systems such as the Magellanic Clouds.

\begin{acknowledgments}

The \textit{Fermi} LAT Collaboration acknowledges generous ongoing support
from a number of agencies and institutes that have supported both the
development and the operation of the LAT as well as scientific data analysis.
These include the National Aeronautics and Space Administration and the
Department of Energy in the United States, the Commissariat \`a l'Energie Atomique
and the Centre National de la Recherche Scientifique / Institut National de Physique
Nucl\'eaire et de Physique des Particules in France, the Agenzia Spaziale Italiana
and the Istituto Nazionale di Fisica Nucleare in Italy, the Ministry of Education,
Culture, Sports, Science and Technology (MEXT), High Energy Accelerator Research
Organization (KEK) and Japan Aerospace Exploration Agency (JAXA) in Japan, and
the K.~A.~Wallenberg Foundation, the Swedish Research Council and the
Swedish National Space Board in Sweden.
 
Additional support for science analysis during the operations phase is gratefully acknowledged from the Istituto Nazionale 
di Astrofisica in Italy and the Centre National d'\'Etudes Spatiales in France.

Resources supporting this work were provided by the NASA High-End Computing (HEC) Program through the NASA Advanced Supercomputing (NAS) Division at Ames Research Center.

\end{acknowledgments}


\bibliography{SMC_DM}

\end{document}